\documentclass[amsmath,amssymb,aps,pra,]{revtex4-1}
\usepackage{amsmath}
\usepackage{blkarray}
\usepackage{braket}
\usepackage{graphicx}
\usepackage{bm}
\usepackage[style=base]{caption}
\usepackage{subcaption}
\usepackage{stackengine}
\begin{document} 

\preprint{APS/123-QED}

\title{Quantum correlations in a  cluster  spin model with three-spin interactions}
\author{Sadaf F}
\email{20phph18@uohyd.ac.in}

\author{V. Subrahmanyam}
\email{vmanisp@uohyd.ac.in (corresponding author)}
\affiliation{School of Physics, University of Hyderabad, Gachibowli, Hyderabad 500046, India}

\date{\today}


\begin{abstract}

An exactly solvable cluster spin model with three-spin interaction couplings $J_x$  (for XZX spin components)  and  $J_y$ (for YZY spin components)  in the presence of a transverse magnetic field $h$  for a spin chain is investigated. For $h=0$, and with only one nonzero interaction strength, the ground state is the cluster state.  Through the Jordan-Wigner fermion mapping, the odd sites and the even sites form two separate  transverse-XY chains, connected only through the boundary terms.  Consequently,  all measures of quantum correlations  for nearest neighbour spins, the  concurrence, the quantum mutual information and the quantum discord are all zero in the ground state.
The dynamics is spin conserving  for $J_y=J_x$,  exhibiting a line of critical points for $|h/J_x|\le 2$, with an uncorrelated direct product ground state for $|h/J_x|>2$. There are several  quantum critical  points in the parameter space, with multi-fold degenerate ground states. The magnetisation and the global entanglement measure exhibit strong singular features for the spin conserving case. The next-neighbour quantum correlation measures are investigated analytically,  which exhibit singular features in the vicinity of degeneracy critical points. 
The $J_y$- and $h$- derivatives of the concurrence  exhibit singular peak  behaviour near the degeneracy critical points, except in the spin conserving case where the derivatives are zero. 
 
\end{abstract}
\maketitle

\section{Introduction}

Quantum entanglement, that has no classical counterpart, has been studied extensively in the last two decades in various physical systems\cite{Horod}, in particular in condensed matter many-body systems, viz. in quantum spin chains, the strongly correlated electron-spin systems\cite{vs, Amico,apvs}. It is recognised as resource in quantum information science,  viz. it has been successfully used for carrying out quantum computation tasks\cite{Ekert, Gott, Bennett, Maika}, for teleporting quantum states\cite{Bennett1},  and for quantum cryptography\cite{Ekert1}.
 Over the last few years,  many measures of quantum entanglement and correlations have been proposed and studied,  such as the concurrence measure, the quantum discord, the global entanglement, the block entanglement entropy\cite{MB, woot}. These measures have been investigated for the ground state of various  spin chain models\cite{arn,chen,shi},  and to study  quantum phase transitions \cite{osborn, osterloh, vimalesh}. The behaviour of entanglement measures is seen to be quite different in the vicinity of a quantum critical point in spin chains.
More recently,  quantum coherence and entanglement in many-body systems \cite{buch, liu}, and non-locality and entanglement in spin systems \cite{niezgoda} have been investigated. 
The dynamics of  quantum correlation and information measures for various spin chain models have been extensively studied over the last few years. The anisotropic Heisenberg-XY spin chains and quenched 1-dimensional spin systems have been explored \cite{vs,ZMwang} from the view point of quantum correlations and entanglement. The quantum correlations in the spin chain dynamics have been investigated \cite{ss1, vimalesh2}. The propagation of quantum correlations and  information scrambling have been investigated using the out-of-time ordered correlators \cite{otoc}, and using the tripartite mutual information \cite{ss2}. 

 Local pairwise quantum correlations depend on the diagonal and off-diagonal correlation functions, that are traditionally studied,  in a complicated and unintuitive way, 
the off-diagonal correlation function should dominate over the diagonal correlation function for a nonzero pair quantum correlation measure.
Even if the concurrence measure of  entanglement is zero for a joint two-qubit state,  quantum correlations can still exist with the quantum discord being nonzero,  as seen
in many  systems\cite{Ollivier, Raoul, kundu}.  The main feature  of the pair entanglement measure in  spin chains is that only nearest neighbour pair concurrence is nonzero,  exhibiting singular behaviour in the vicinity of quantum critical points\cite{osborn, osterloh,vimalesh}. For studying the correlations that are global in nature, there are two popular measures, the global entanglement and the block entanglement entropy, that have been studied extensively. The global entanglement measure \cite{Meyer, Brennen} uses the averaged single qubit entanglement with the rest of the chain, and it is seen to display stronger correlations than the pair concurrence and the quantum discord in Kitaev-type spin chains\cite{vimalesh}.  
 
 Here, we will construct and investigate, from the view of quantum correlations and entanglement,  a cluster spin model with three-spin interactions (of XZX and YZY spin components) along with a transverse magnetic field. In the cluster spin model, the nearest-neighbour quantum correlation measures are zero, but  nonzero measures of quantum correlations exist  for the next neighbour pair of spins. This is in contrast with spin chain models with pairwise interactions, viz. the transverse-XY model that exhibit nonzero quantum correlations only for  nearest neighbour spins . In the limit of only one type of three-spin interaction being nonzero in the absence of the magnetic field, the ground state is the cluster state\cite{Shor}, a highly entangled state, that has been used for measurement-based quantum computing\cite{briegel}. It has been proposed that multi-spin interactions can be experimentally realised in optical lattices of ultra cold atoms\cite{pachos}. Three-spin interactions along with usual two-spin interactions in cluster spin chains have been investigated over last two decades,  for phase transitions and localisable entanglement\cite{skrov, son, smacchia}, for the topological majorana correlations\cite{ohta}, for quantum criticality and symmetry-protected cluster states\cite{lahtinen}, for quantum coherence and localisation\cite{bahri}. The three-body interactions have been investigated for superconducting qubits\cite{menke}, and in quantum simulations in weakly driven quantum systems \cite{petiziol}. The competing many-body interactions in trapped ion systems \cite{katz}, and higher-dimensional magneto-electric order in  spin chains  with 3-spin interactions\cite{durganandini} have been investigated. It is interesting to investigate the behaviour of the quantum correlation and information measures in spin chains with three-body interactions.
  
In Section II, we present the exact solution of a general cluster model with XZX and YZY three-spin interactions along with a magnetic field along the z-direction, using the Jordan-Wigner transformation that maps the spin system to spin less fermions. We will see that, in the fermion picture, in the cluster model the even sites and odd sites form separate transverse-XY spin chains, coupled only through boundary terms.
This implies that the nearest neighbour pairs do not have correlations. 
We will see  that there are degeneracy critical points in the parameter space, where the ground state is multi-fold degenerate. In Section III, we explore the single-site and two-site reduced density matrices, and find the magnetisation and diagonal and off-diagonal correlation functions analytically. A detailed analysis of the next nearest neighbour concurrence, and its derivatives with respect to the interaction strength and the magnetic filed is  presented in Section IV. 
In Section V, we will explore analytically the pairwise quantum mutual information, followed by the conditional entropy and the quantum discord in Section VI.  We will discuss the global entanglement measure in Section VII, that exhibits singular behaviour for the spin conserving dynamics. The conclusions are presented in Section VIII.

\section{ A Cluster Spin Model and its Exact Ground State }

In this paper, we will construct a general cluster spin chain model with three-spin interactions and study its exact solution through the Jordan-Wigner transformation.  We will investigate the measures of quantum correlations for the ground state in the following sections.
We consider a  one-dimensional spin chain of $N$ spins with three-spin interactions. 
For a set of three neighbouring spins, there are three-spin interactions XZX and YZY of the respective spin components, i.e. the nearest neighbour pairs (the first and the second, the second and the third) interact through X-Z  and  Y-Z  spin components, and the next neighbour spins (the first and the third) interact with X-X and Y-Y interactions of spin components. That is, the cluster spin model has three-spin interactions,  as a straightforward extension of the transverse-XY model that has two-spin  interactions  of X-X and Y-Y spin components for nearest neighbour spins. Thus, there are in-built correlations between nearest-neighbour spins and the next neighbour spins in  the cluster model  with three-spin interactions.  As we will see below, though the three-spin interactions  may seem complicated, the cluster spin model can be exactly diagonalised for all eigenstates, similar to the case of the transverse-XY model. 
 
For a one-dimensional spin-1/2 chain, the cluster spin model Hamiltonian is given by, 
\begin{equation}
H= -J_{x} \sum_{l=1}^{N} \sigma_{l-1}^x \sigma_{l}^z \sigma_{l+1}^x - J_{y} \sum_{l=1}^N \sigma_{l-1}^y \sigma_{l}^z \sigma_{l+1}^y - h\sum_{l=1}^N \sigma^z_{l}.
\label{H1:ham1} \end{equation}
The spin operators $ (\sigma^x_{l},\sigma^y_{l},\sigma^z_{l})$ represents $x$, $y$ and $z$ components of Pauli spin operators at each site $l$. 
The three-spin interaction  strengths $J_{x}$ and $J_{y}$ correspond  to the $XZX$ and $YZY$ spin components of the respective  contiguous thee-spin clusters, and $h$ is the external transverse magnetic field.  We employ periodic boundary conditions to gain the  translational invariance.
The Floquet dynamics of the above cluster model  with only one interaction strength (i.e, $J_y=0$) has been studied\cite{Verga}, showing a dynamical transition from low to high-entanglement phase.  There are investigations of related spin chain models with three-spin interactions for potential experimental realisation,  phase transitions, topological correlations, criticality and localisation [27-33].

The above Hamiltonian has the cluster state, a highly entangled cluster state,  as the ground state if one of the interaction strengths is zero, in the absence of the magnetic field\cite{Shor, Raussendorf}. 
Traditionally the cluster state is written in terms of $\sigma^x$ eigenstates and $\sigma^z$ operators of neighbouring spins, but here we will write in terms of $\sigma^z$ eigenstates and $\sigma^y$ operators of neighbouring spins.
Let us define a controlled $Y$ operator  for two spins at sites $i$ and $i+1$,  as  ${\cal Y}_i= 1- (1-\sigma_i^y)(1-\sigma_{i+1}^y)/2$. The cluster state is given by
\begin{equation} | \psi_{cluster}\rangle = \prod_i {\cal Y}_i |\uparrow_1\uparrow_2..\uparrow_N\rangle,
\end{equation}
where $|\uparrow_l\rangle$ is the eigenstate of $\sigma_l^z$ with eigenvalue 1.  The cluster state is the ground state of the above Hamiltonian (when $J_x=0=h$) with eigenvalue $-J_yN$, as each term in the second sum of Eq.1 will give an eigenvalue 1 on acting on the cluster state.  In this case, the Hamiltonian is a sum of stabiliser operators\cite{Gottesman}, important for quantum error correcting codes, that commute with each other, giving $N$ conserved quantities. The cluster state is a highly entangled state with the Schmidt dimension growing exponentially with $N$. Any pair of spins can be projected onto a maximally entangled Bell pair state, by measuring the other spins in an appropriate measurement basis. Thus, the cluster states are in important information resource in measurement based quantum computation. Similarly, when $J_y=0,h=0$, the ground state is a cluster state, using the controlled $\cal X$ operator by replacing the $\sigma^y$ by $\sigma^x$ in $\cal Y$. As the parameters are varied,  the ground state of the cluster spin model can display a variety of quantum correlations, that will be explored in detail in the next sections.

The cluster model Hamiltonian  can be diagonalised through a series of transformations, similar to the transverse-XY model\cite{Lieb-Mattis}. Firstly, we employ the Jordan-Wigner transformation to convert spin operators into fermion creation and annihilation operators at every site, given by
\begin{equation}
\sigma_{l}^{+} =  e^{i\pi \sum_{m=1}^{l-1} c_{m}^\dagger c_{m}}   c_{l}^\dagger,  
~~\sigma_{l}^z = 2 c_{l}^\dagger c_{l} - 1.
\end{equation}
Here, fermion operators $c_{l}$ and $c_{l}^\dagger $ can annihilate and create a fermion at site $l$, corresponding to the spin lowering and raising spin operators. In fermion space,
$n_{l}= c_{l}^\dagger c_{l}$ is a number operator that counts the number of fermions  at site $l$.
The spin basis states $|\uparrow\rangle, |\downarrow\rangle$ for site $l$  correspond to  the fermion occupied Fock state $|n_{l}=1\rangle$ and the unoccupied Fock state $|n_{l}=0\rangle$.
The cluster spin Hamiltonian given above  does not conserve the number of up spins in general (except for the spin conserving case of $J_y=J_x$), as each of the interaction terms can increase or decrease the number of up spins by two, i.e. the dynamics decouples sectors with even and odd number of up spins. 
This implies that,  the Hamiltonian can be diagonalised separately in the sectors with an even number or odd number of up spins, analogously with the transverse-XY model.
This translates to the two sectors for the fermion number,  even or  odd $N_F$. The periodic boundary condition for the spin model translates to the  periodic boundary condition for fermion operators for the odd $N_F$ sector, i.e. $c_l=c_{N+l}$. For the even $N_F$ sector, we have anti-periodic boundary condition for the fermion operators, i.e. $c_l=-c_{N+l}$.  In terms of the fermion operators, it is straightforward to see that the interaction term with a given $l$ in Eq.1, will involve operators of sites $l-1$ and $l+1$. The intervening third spin in the three-spin cluster does not appear in the fermion picture, exhibiting only next-neighbour interactions. That is, the cluster spin model splits into two transverse-XY spin chains containing even and odd sites separately. However, the boundary interaction terms (corresponding to terms with  $l=1$ and $l=N$ in Eq.1) involve the total number of fermions in the system, thus coupling the two chains. Using the convention of periodic (anti-periodic) for   $N_F$ odd (even), the boundary terms become similar to the bulk interaction terms.  The Hamiltonian, thus,  takes the form,
\begin{equation}
H=\sum_{l=1}^{N} J_x (c_{l-1}^\dag-c_{l-1})(c_{l+1}^\dag+c_{l+1}) +J_y (c_{l-1}^\dag-c_{l-1})(c_{l+1}^\dag+c_{l+1}) - h \sum_{l=1}^N (2c_l^\dag c_l-1),
\end{equation}
exhibiting a coupling for next-neighbour spins only. The eigenstates of the cluster model will be direct products of even-site states and odd-site states. The global constraint that $N_F$ is even, for example, will lead to a superposition of direct products. That is,  the even sites and the odd sites can have either even and even, or odd and odd number of fermions, when the total number of fermions is even. Let us denote the total chain as consisting of two parts $A$ (of even sites) and $B$ (of odd sites). Now,  an even-sector  ($N_F$ even) eigenstate for the chain will be a superposition of even-even and odd-odd states for the two parts, i.e. an entangled state of the form  $\alpha |even\rangle_A |even\rangle_B +\beta |odd \rangle_A |odd\rangle_B$. However, the amount of entanglement between the even and odd chains is independent of $N$.

Exploiting the translational invariance of the spin chain, we transform the position space fermion operator to the momentum space fermion operator, through
\begin{equation}
c_{l}= \frac{e^{\frac{i\pi}{4}}}{\sqrt{N}} \sum_{k} e^{ikl} c_{k}.
\end{equation}
The allowed values for $k$ are different for even and odd $N_F$ sectors, and whether the total number of sites $N$ is even or odd. We consider the case of  even $N$ only here.  Thus, we have for the even $N_F$ sector, $k = \frac{n\pi}{N} \hspace{4pt} \lbrace n = \pm1,\pm2,...,\pm(N-1)\rbrace$. For odd $N_F$ sector, we have $k = \frac{n\pi}{N} \hspace{4pt} \lbrace n = 0,\pm 2, \pm4,...,\pm(N-2),N\rbrace$. Thus, in the thermodynamic limit,  $k$ varies from -$\pi$ to $\pi$.

In terms of the momentum space fermion operators the Hamiltonian is given by, for the even $N_F$ sector,
\begin{equation}\label{H2:ham2}
H=\sum_{0<k<\pi} A_k(c_{k}^\dagger c_{k} + c_{-k}^\dagger c_{-k})+ B_k (c_{k}^\dagger c_{-k}^\dagger + c_{-k}c_{k}) \\- hN,
\end{equation}
where the sum over $k$ is over the allowed values discussed above for the even sector,  and the $k-$dependent coupling strengths for the bilinear terms above are given by
\begin{equation} A_k=-h+(J_{x}+J_{y})\cos2k,~~~ B_k=(J_{x}-J_{y})\sin2k.
\end{equation}
 Similarly, we can write the fermion Hamiltonian for the odd sector, by adding $k=0$ and $k=\pi$ terms. Now, the Hamiltonian is a direct sum of $N/2$ terms labeled by positive $k$ values, However, for each $k$ term, still there is coupling between the fermion operators  $c_k$ and $c_{-k}$. We can decouple this also, and write the Hamiltonian in a diagonal form as given below. 

We can re-write the above Hamiltonian in a simpler form using two-component spinor operators,  as
\begin{equation}\label{H3:ham3}
H=\sum_{k>0} \psi_{k}^\dagger M_{k} \psi_{k},  ~~{\rm where~~}\psi_{k} = 
\begin{pmatrix}
c_{k}\\
c_{-k}^\dagger\\   
\end{pmatrix},\end{equation}
where  the matrix between the spinor operators is given in terms of Pauli matrices, as  $M_{k} = \omega_k  (\cos{\theta_k}\sigma^z + \sin{\theta_k} \sigma^x)$. Here, the eigenvalue coefficient is
given by $\omega_k=\sqrt{A_k^2+ B_k^2}$, and the angle is defined by $\tan{\theta_k}= -B_k/A_k$. In the above, the Hamiltonian is decoupled for $N/2$ different values the momentum, for $0<k<\pi$, but still the fermion operators corresponding to $k$ and $-k$ are coupled.
The next step is to employ a Bogolyubov transformation to diagonalise the Hamiltonian  for given $k$ value, written in terms of spinor operators.  That is, we go to the diagonal basis of the matrix $M_k$, that results in a new set of fermion operators that are linear combinations of $c_k$ and $c_{-k}^\dagger$.

The unitary matrix for the desired transformation is,
\begin{equation}
U =
\begin{pmatrix}
\cos\theta_k/2 & \sin\theta_k/2 \nonumber \\
\sin\theta_k/2 & -\cos\theta_k/2 \\
\end{pmatrix}
\end{equation}
The columns of this matrix are the eigenvectors of the matrix form of the operator $M_{k}$, corresponding to the eigenvalues $\pm \omega_{k}$ respectively. We have new spinor operators, defined by  $\eta_k= U\psi_k$,  that relates the $\eta$-fermion annihilation operators to the $c$-fermion annihilation and creation operators, we have 
\begin{equation}
\begin{pmatrix}
\eta_{1k} \\
\eta_{2k} \\
\end{pmatrix}
= \begin{pmatrix}
\cos\frac{\theta_k}{2} c_{k} + \sin\frac{\theta_k}{2} c_{-k}^\dagger \\
\sin\frac{\theta_k}{2} c_{k} - \cos\frac{\theta_k}{2} c_{-k}^\dagger\\
\end{pmatrix}.
\end{equation}
Finally, the Hamiltonian has a diagonal form in terms of these $\eta$-fermion operators, given by
\begin{equation}\label{H4:finalham}
H = \sum_{k>0} 2 \omega_{k} (\eta_{1k}^{\dagger} \eta_{1k} - \eta_{2k}^{\dagger} \eta_{2k}).
\end{equation}
For each of $N/2$ values of $k$, there are two independent fermion modes, each $\eta$-fermion mode annihilation operator being a linear combination of $c$-fermion annihilation operator with $k$ and $c$-fermion creation operator with $-k$. For each $k$, the Hamiltonian has four eigenstates,  $\ket{vac}, \eta_{1k}^{\dagger}\ket{vac},\eta_{2k}^{\dagger}\ket{vac},\eta_{1k}^{\dagger}\eta_{2k}^{\dagger}\ket{vac}$, where $|vac \rangle $ is the vacuum state with no occupation of both $\eta$ fermions, with the corresponding eigenvalues
$0, 2\omega_k, -2\omega_k, 0$. The $\eta$-fermion vacuum state is related to the $c$-fermion number states through $|vac\rangle= |0\rangle_k|1\rangle_{-k}$.

The ground state of the Hamiltonian is given by occupying each $k$ state by $\eta_2$ particle, with the ground state energy, $E_{g}= -2\sum_{k} \omega_k$.
Thus, the ground state $|G\rangle$ of the cluster spin model is a direct product over all  $k$ values. It can be written in terms of $\eta_{2}$-fermion states, and in terms of  $c$-fermion states (using Eq.9), we have
\begin{equation}\label{eq1:gs}
\ket{G} = \prod_{k>0} \eta^{\dagger}_{2k} \ket{vac} 
=\prod_{k>0} (\cos\frac{\theta_{k}}{2} - \sin\frac{\theta_{k}}{2} c_{k}^{\dagger} c_{-k}^{\dagger}) \ket{00..0}.
\end{equation}
Thus, the ground state is a coherent superposition of all even-numbered $c$-fermion states, similar to the transverse XY model. However, the coefficient for the fermion number states holds the key to the structure of the state  and the distribution of quantum correlation measures as we shall discuss below.  The probability amplitude for the occupied $k$ states is given by,
\begin{equation}
\sin{\theta_k\over 2}
= {1\over \sqrt{2}}\sqrt{1- {A_k\over \omega_k}}.
\end{equation}
Thus, we have the cluster spin model ground state, constructed as a direct product  of $k$-dependent states with probability amplitudes corresponding to $k$ and $-k$ of $c$-fermion Fock states either both occupied or unoccupied. Though it a product state in the $k$ space, it can display a complicated quantum correlations and entanglement structure in the position space, as we will discuss in the following sections.

The structure of the ground state for the spin conserving case,  $J_y=J_x$,  is  different from other cases.  For this case,  we have $B_k=0$ from Eq.7, implying that there are no spin non-conserving terms in the Hamiltonian in Eq.6.  Thus,  the mode energy is $\omega_k=|A_k| $.  Now, for $A_k>0$, the corresponding $k-$states are unoccupied as $\sin{\theta_k/2}$ vanishes, and for $A_k<0$ the corresponding $k-$states are occupied with probability one. This implies, from Eq.7, that for $|{h/ J_x}|>2$, the ground state is a direct product state, either all momentum states occupied or unoccupied, corresponding to all up-spin state or all down-spin state, i.e. the ground state is an uncorrelated spin state. It is quite interesting  to see that the three-spin interactions and the transverse magnetic field conspire to cancel the quantum correlations, and give in an uncorrelated direct product spin state  as the ground state in the spin conserving case for $|h/J_x|>2 $.
 For other values of the magnetic field,  $|h/J_x|<2$, there is a line of quantum critical points, and the ground state is not a direct product state that may have quantum correlations.  Moreover, for some value of $k$ the coefficient $A_k$ can vanish  depending on the magnetic field, implying that the mode energy is zero, leading to a degeneracy in the ground state. We will discuss the ground state degeneracy and the degeneracy critical points below for all parameters.

As we have constructed the ground state by occupying $\eta_2$ particles for all $k$, a unique ground state is obtained if the mode energy is nonzero $\omega_k\ne0$ for all $k$. A degeneracy of four will arise if the mode energy vanishes for a given $k$.  For example, if $\omega_{k_1}=0$,  then a ground state will have $\eta_2$ particle occupying all other $k$ values, but for $k=k_1$ we can take either of the unoccupied state,  the single $\eta_1-$occupied state, the single $\eta_2-$occupied state, the doubly occupied state, since all four states have zero energy.  Similarly, if the mode energy vanishes for $l$ different values of $k$,  we will have $4^l$-fold degenerate ground states. Thus, we have a gapless phase with 4-fold or more degenerate ground state near the degenerate critical points, and a gapped phase with a unique ground state away from the critical points. We expect the distribution of the quantum entanglement and quantum correlations to exhibit a different structure in the gapless phase, as we shall explore in the following sections.

The mode energy, $\omega_k=\sqrt{A_k^2+B_k^2}$ depends on the interaction strengths and the magnetic field, through the $k-$dependent coupling strengths shown in Eq.6. For the mode energy to vanish, we need both $A_k$ and $B_k$ to vanish for a given value of $k$ that depends on the coupling strengths $J_x, J_y$ and $h$.
There are three different cases here:\\
\noindent \underline {For the  spin conserving case of  $J_x=J_y:$}  We  already have $B_k=0$ for all $k$. The other coefficient $A_k$ vanishes for a $k$ value satisfying
$\cos 2k= h/2J_x$. This implies a four-fold degeneracy in the ground state for an every value of the magnetic field satisfying  $|h/J_x|<2$.  This implies a line of degeneracy (critical) points in the spin conserving phase.

\noindent \underline {For  $J_x=-J_y:$} In this  case, $A_k$ will vanish only for zero magnetic field. The other coefficient $B_k$ will vanish for $k=\pm \pi/2$. For a thermodynamic system, it will vanish for $k=0, \pi$ also.  This implies an isolated degeneracy point for $h=0$. 

\noindent \underline {For $|J_y/J_x| \ne 1:$} In this case, $B_k$ will vanish only for $k=0$ and $k=\pi$, which are allowed values in the thermodynamic limit. For these $k$ values, the  coefficient $A_k$ will vanish for a magnetic field satisfying $h=J_x+J_y$, leading to a degenerate ground state for a  large size. Additionally, $B_k$ vanishes for $k=\pm \pi/2$ for all sizes with $N$ even. This implies a degeneracy point for $-h=J_x+J_y$.
Thus,  for large system sizes, the degeneracy critical points exist for 
$J_x+J_y = \pm {h}.$

Thus, there are two different types of degeneracy critical points. A line of quantum critical points for the spin conserving case, $J_y=J_x$, till a cutoff value of the magnetic field, $|h/J_x|=2$. Beyond the cutoff, the ground state is uncorrelated. For the spin non-conserving case, there are isolated quantum critical points, as discussed above. We expect that the behaviour of measures of the quantum correlations and entanglement to be different in the vicinity of degeneracy critical point, that will be discussed in the following sections.

\begin{figure*}[t]
\includegraphics[scale=0.4]{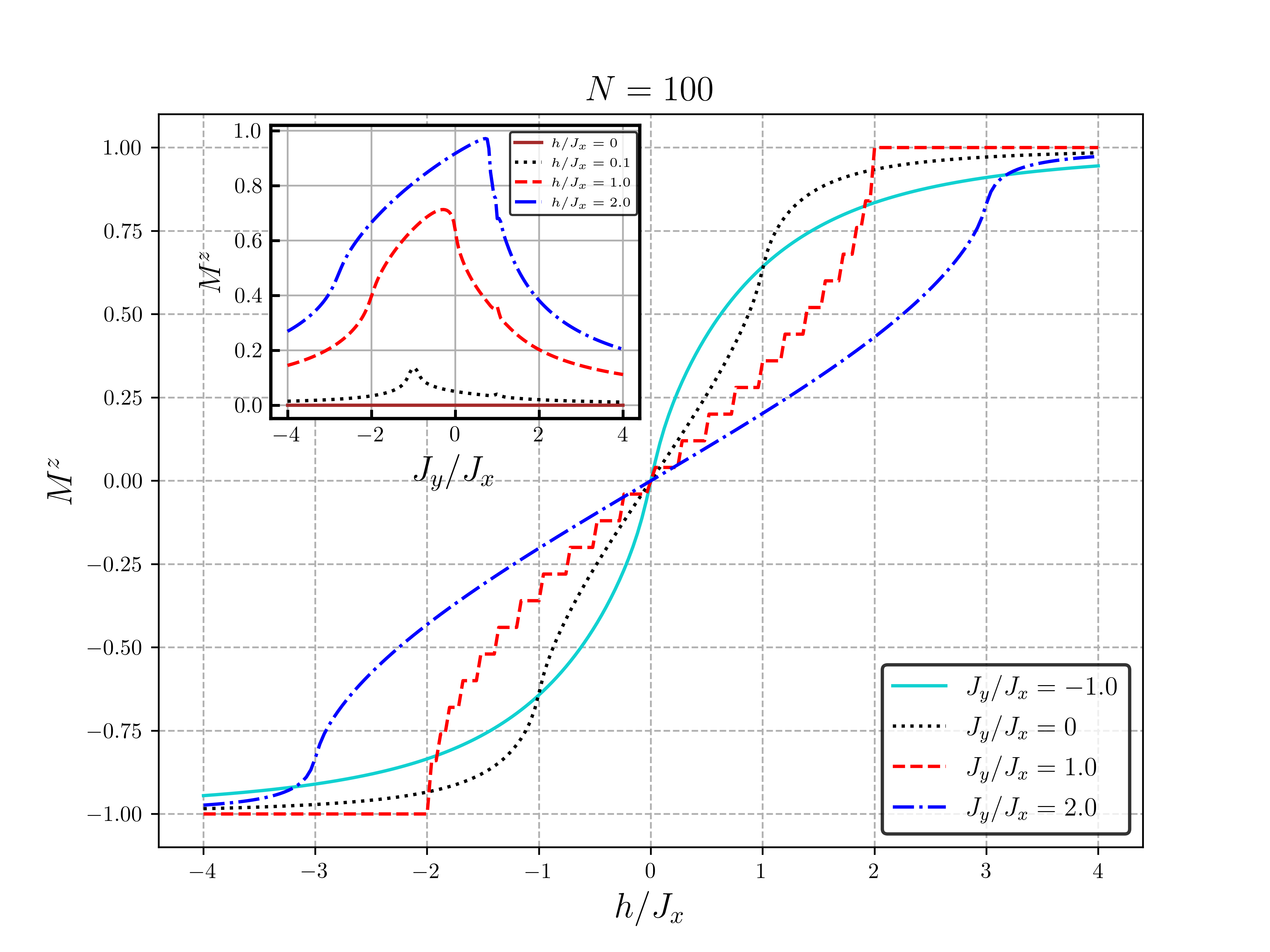}


\caption{The magnetisation $M^z$ as function of the transverse magnetic field $h$, for different ratios of the coupling strengths $J_y/J_x$  in the ground state for $N=100$. In the spin conserving case, $J_y=J_x$,  a staircase structure can be seen, due to a line of degeneracy critical points.  $|M^z|$ saturates to the maximum value for
$|h/J_x|>2$, as the ground state is a direct product state in this region. The inset shows the $M^z$ as a function of the ratio $J_y/J_x$, for different values of $h$.  Smooth peaks can be seen close to the degeneracy points, with a small blip at $J_y=J_x$. }
\end{figure*}

\section{Magnetisation and two-spin correlation functions}

In order to explore quantum correlations and entanglement, we need the matrix elements of single-site and two-site operators in the ground state. 
In general, the state of a sub-system of spins,  obtained by a partial trace over the remaining spins, of a pure many-qubit pure state will be a mixed state.  That is the subsystem is represented by a reduced density matrix.  For the ground state, the single-site reduced density matrix $\rho_l$ at the site $l$  is defined as $ \rho_{l}= Tr^\prime\ket{G}\bra{G}$, where the prime over trace indicates a partial trace over all  spins excluding the spin at the site $l$.
Similarly, the two-site reduced density matrix for sites $l$ and $m$ is given by a partial sum over all spins except for the marked pair of spins,  $\rho_{lm}= Tr^{\prime}\ket{G}\bra{G}$. The local uniform interactions between the three spin clusters,  and the periodic boundary conditions result in the single-qubit density matrix being the same for all sites i.e. $\rho_1=\rho_{l}$ for all $l$ . Similarly, there is only one independent two-site reduced density matrix for two spins with a given distance between them, i.e., $\rho_{12}=\rho_{23}=\rho_{l,l+1}$,  and $\rho_{13}=\rho_{24}=\rho_{l,l+2}$.  A state with an even number of fermions does not mix with an odd-number state, many of the off-diagonal matrix elements of the reduced density matrix will be zero. This indicates  some specific off-diagonal elements are non-zero of the reduced density matrix that correspond to even-even or odd-odd connections.

Using the $\sigma^z_l$ diagonal basis represented by $\ket{0}$ and $\ket{1}$  for the spin at site $l$, the single-qubit reduced density matrix can be written as,
\begin{equation} \label{eq2:rho1}
\rho_{l}=
\begin{pmatrix}
\langle  {1+\sigma^z_{l}\over 2} \rangle & 0 \\

 0 &  \langle { 1-\sigma^z_l \over 2} \rangle \\ 
\end{pmatrix}.
\end{equation}
Here, the angular brackets denote an expectation value in the ground state. The off-diagonal matrix elements, related to $\langle \sigma^+_l\rangle$, are zero as the ground state is an even-number state.
The diagonal matrix element $\langle (1+\sigma^z_{l})/2\rangle$ corresponds to the ground state expectation value of the $l-$site number operator in the fermion language, $n_{l}= \left\langle c_{l}^{\dagger}c_{l} \right\rangle$. 
It is straightforward to calculated $n_l$ using the momentum space operators,  we have
\begin{equation} \label{eq3:NO}
n_{l}= \left\langle c_{l}^{\dagger}c_{l} \right\rangle = \frac{2}{N} \sum_{k>0} \sin^2 \frac{\theta_{k}}{2} .
\end{equation}
As argued earlier, the number operator expectation value does not depend on the site index.
The magnetisation, which is an expectation value of total spin in $z$ direction, can be written in terms of the number operator calculated above as,

\begin{equation}
M^z= \frac{1}{N}\,\,\sum_l \left\langle \sigma_l^z \right\rangle = 2n_l-1.
\end{equation}
The behaviour of magnetisation $M^z$ in the ground state  depends on the magnetic field $h$ and coupling strength $J_x$ and $J_y$, as shown in Fig.1 for a finite size $N=100$.
As a function of the magnetic field $h$, the magnetisation varies  smoothly from $-1$ to 1 if there is no degeneracy region, for a fixed value of $J_y/J_x$.  For $J_y/J_x=0, 2$, $M^z$ exhibit a smooth behaviour as function of $h$. 
For the spin conserving case of $J_y=J_x$,  there is a line of degeneracy points between $h/J_x=-2$ to $h/J_x=2$, we  can see a jagged structure (a staircase pattern).   In this case, for $h/J_x>2$ the ground state is a direct product state, either all spin up (for $h>0$) or down, the magnetisation saturates with $|M^z|=1$ in this range. For $J_y=-J_x$, a degeneracy point exists for $h=0$, the magnetisation shows infinite slope singular behaviour near zero magnetic field.
The inset shows the magnetisation plotted as a function of the ratio of interaction strengths $J_y/J_x$ for different values of the magnetic field $h$. For $h=0$, the magnetisation is zero for all values of the coupling strengths. For a nonzero $h$, the magnetisation exhibits a peak as the ratio $J_y/J_x$ is increased, with the peak shifting to higher values as $h$ is increased.  The signature of the degeneracy critical point can be seen as a blip at $J_y/J_x=1, -1$.

For  investigating the local pair  quantum correlations, namely the concurrence measure of the entanglement and the quantum mutual information, we need to analyse the diagonal and off-diagonal matrix elements of the two-qubit reduced density matrix. We construct  the two-qubit reduced density matrix $\rho_{lm}$ for spins at sites $l$ and $m$,  using the $\sigma^{z}$ diagonal basis states $\ket{00},\ket{01},\ket{10} and \ket{11}$ for the two sites. The reduced density matrix  has  a X-form, given by
\begin{equation}
\rho_{lm} = 
\begin{pmatrix}
u_{lm} & 0 &0 & x_{lm}\\
0& w_{lm}& z_{lm}&0 \\
0& z_{lm}^{*} & w_{lm} & 0\\
x_{lm}^{*} & 0 & 0 & v_{lm}\\   
\end{pmatrix}.
\end{equation}
We need to compute only two off-diagonal matrix elements  $x_{lm}$ and $z_{lm}$, for the two-site reduced density matrix above, as odd-even and even-odd connections are zero.
The diagonal and off-diagonal matrix elements represent  the respective spin-spin correlation functions. The off-diagonal matrix element  $x_{lm}$ is non-zero only for the spin non-conserving case $J_y\ne J_x$,  as the Hamiltonian does not commute with the total z-component of the spin in this case. In our case, the second and third diagonal matrix elements are equal due to the translational invariance, but for a general X-state they may not be equal.
We can express the above matrix elements as the ground state expectation values of  spin operators, further using the Jordan-Wigner transformation we can express them in terms of fermion operators, we have,
\begin{equation}\label{eq5:RDM}
\begin{aligned}
u_{lm} &= \left\langle \frac{1+\sigma_{l}^{z}}{2} \frac{1+\sigma_{m}^{z}}{2}  \right\rangle 
= \left\langle c^{\dagger}_{l} c_{l}c_{m}^{\dagger}c_{m} \right\rangle , ~~ v_{lm}=1-u_{lm}-2w_{lm},      \\
w_{lm} &= \left\langle  \frac{1+\sigma_{l}^{z}}{2}\frac{1-\sigma_{m}^{z}}{2}  \right\rangle = \left\langle c_{l}^{\dagger}c_{l} \left( 1-c_{m}^{\dagger}c_{m}\right)  \right\rangle, \\
z_{lm} &= \left\langle \sigma_{l}^{+} \sigma_{m}^{-}\right\rangle =\left\langle c_{l}^{\dagger}c_{m}\prod_{n=l+1}
^{m-1} \left( 1-2 c_{n}c_{n}^{\dagger}\right) \right\rangle,
~~x_{lm} = \left\langle \sigma_{l}^{+}\sigma_{m}^{+}\right\rangle =\left\langle c_{l}^{\dagger}c_{m}^{\dagger}\prod_{n=l+1}^{m-1} \left( 1-2 c_{n}c_{n}^{\dagger}\right) \right\rangle .
\end{aligned}
\end{equation}
Let us define auxiliary functions involving  sums over functions of the the mode energies and the other coefficients, given by
\begin{align}
\gamma(\textbf{\textit{p}})&= \frac{2}{N} \sum_{k>0} \cos(pk) \sin^2{\theta_k\over 2},~~
\xi(\textbf{\textit{p}})=-\frac{1}{N} \sum_{k>0} \sin(pk) \sin {\theta_k}.
\end{align}
We can see from the above and Eq. 14. that $\gamma(0)=n_l$.
The nearest-neighbour pair correlation functions can be calculated in the ground state as,
\begin{equation}
z_{l, l+1} =\gamma(1),~
x_{l, l+1} = \xi(1).
\end{equation}
\begin{figure*}[t]
\includegraphics[scale=0.8]{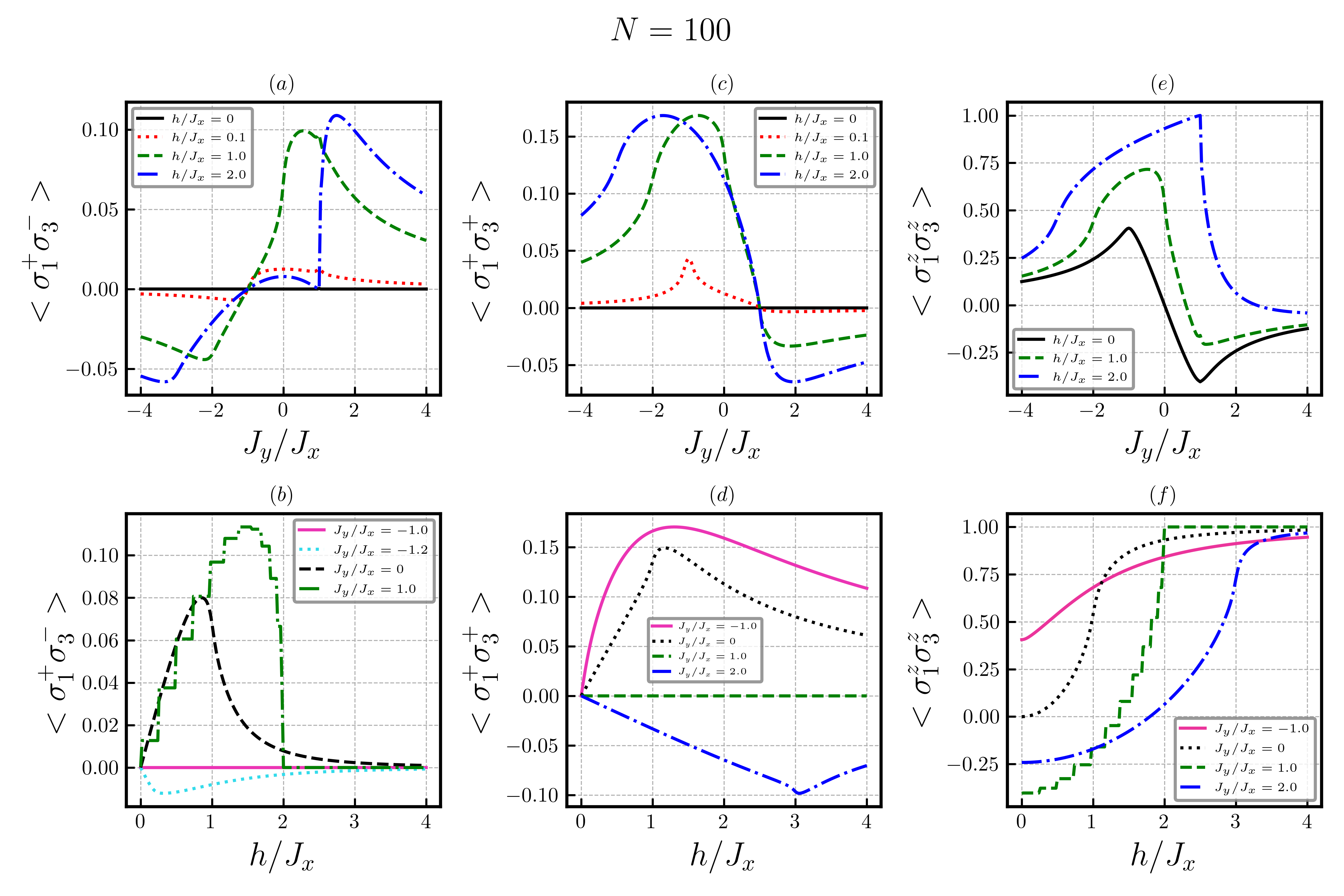}
\caption{ The off-diagonal correlation functions and the diagonal correlation function, shown as functions of $J_y/J_x$ (top), and as functions of $h/J_x$ (bottom) for the next neighbour pair of spins, show smooth behaviour for most of the parameter values. The correlation functions exhibit smooth peaks in the vicinity of degeneracy points. For the spin conserving case of $J_y=J_x$,   $\langle \sigma_1^+ \sigma_3^+\rangle $ is zero as expected, $\langle \sigma_1^+\sigma_3^-\rangle$ shows a stair case structure as a function of the magnetic field, with a cutoff value $h=2J_x$,  as the ground state is a direct product state.  The diagonal correlation function exhibits a staircase structure, with saturation beyond the cutoff.  There is no cutoff value of $h$ for spin non-conserving case.
 }
\end{figure*}The next neighbour off-diagonal correlation functions, that involve four fermion operators, are more difficult to calculate, we have
\begin{equation}
z_{l,l+2}
=  \gamma(2) (1- 2\,n_l)  + 2\,\gamma(1)^2, ~~x_{l,l+2} =\xi(2) (1- 2 n_l).
\end{equation}
Using the diagonal and off-diagonal matrix elements of the reduced density matrix that we have calculated above, we can compute different measures of quantum correlations and entanglement for the next neighbours, as detailed in the following sections.
\begin{figure*}[t]
\includegraphics[width=\textwidth]{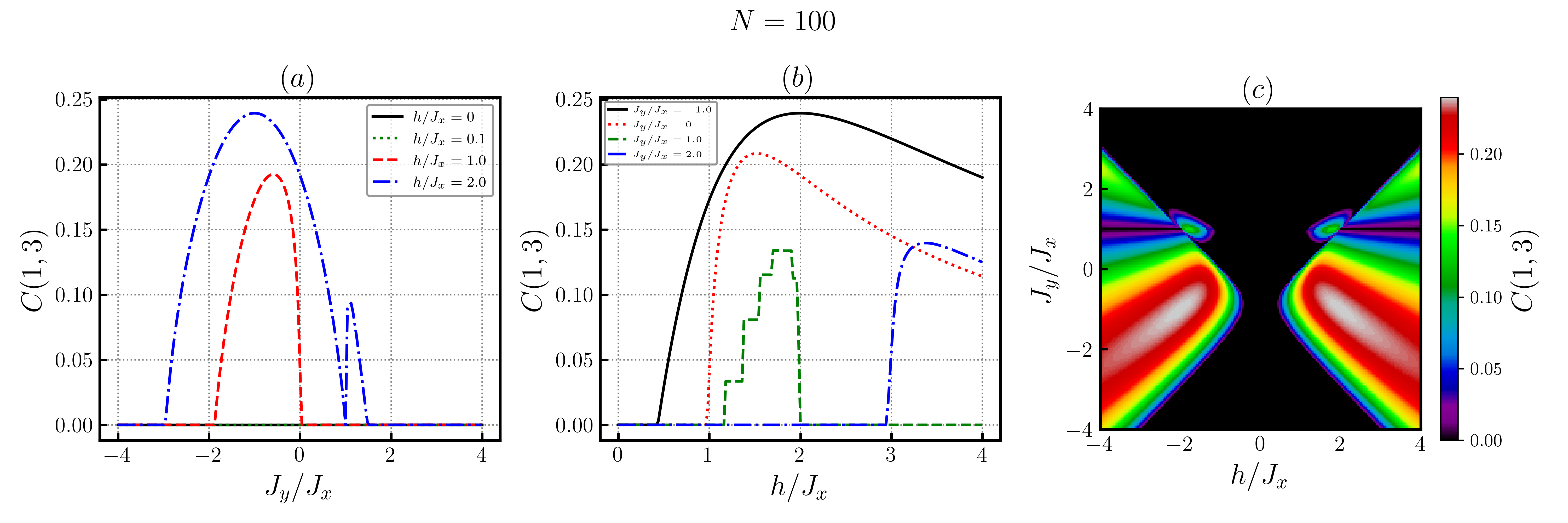}
\caption{
The  next-nearest neighbour concurrence $C(1,3)$ for $N=100$. (a)  It is shown as a function of $J_y/J_x$ in  for different values of the magnetic field.
The concurrence is zero if the magnetic field is close to zero. For $h=2J_x$, the concurrence shows a dip for the spin conserving  case $J_y=J_x$.
 (b) The concurrence is shown as a function of $h/J_x$  for different
values of $J_y/J_x$.  The concurrence becomes nonzero only after a certain value of $h$,  and shows a maximum close to the degeneracy point. For the spin conserving case, it shows a staircase structure, with a cutoff at $h=2J_x$, as the ground state becomes uncorrelated beyond the cutoff value of the magnetic field. (c) The density plot of $C(1,3)$  as a function of $J_y/J_x$ and $h/J_x$. It is zero for $|h/J_x|\le1$, and takes larger values for  $J_y/J_x<0$.
}

\end{figure*}

\section{Concurrence} \label{section: concurrence}
There are several measures to quantify local  quantum correlations and the entanglement for two-qubit mixed states. The  concurrence measure of the entanglement, which was initially introduced as a supplementary tool to calculate the entanglement of formation in the case of two-qubit states \cite{woot}, is widely investigated for spin chains. The concurrence measure can used to  quantify and characterise the non-classical correlation between two qubits in a composite quantum system. For a pair of spins at sites $l$ and $m$, the concurrence $C(l,m)$ ranges from zero, indicating that there is no entanglement (i.e. the joint state is a separable state), to a  maximum value of 1, indicating a maximally entanglement  state (i.e. the joint state is a Bell state).
In this section, our focus is on examining the concurrence $C(1,3)$ as a measure of the entanglement for next-neighbour qubits,  as concurrence $C(1,2)$  is zero for the nearest-neighbour pairs  in the cluster spin model ground state, as detailed in the previous section.

The concurrence for any two-qubits in a general mixed states\cite{woot} is given by,
\begin{equation}
C(l,m) = 2~ {\rm max}(0,\lambda_{1}-\lambda_2-\lambda_3-\lambda_4).
\end{equation} 
Here, $\lambda_{i}$ are the eigenvalues of the Hermitian matrix defined as $\sqrt{\sqrt{\rho_{lm}}\tilde{\rho_{lm}}\sqrt{\rho_{lm}}}$ in descending order, where $\tilde{\rho}$ is the time-reversed density operator,  $\tilde \rho_{lm}=(\sigma_l^{y}\sigma_m^{y})\rho_{lm}^{*}(\sigma_m^{y}\sigma_l^{y})$.
Using the diagonal and off-diagonal matrix elements of two-site reduced density matrix  $\rho_{12}$ for nearest neighbours, and $\rho_{13}$  for next-nearest neighbours, discussed in the previous section, the concurrence can be determined, we have
\begin{equation}
C(l,m) = 2 \hspace{2pt} {\rm max} (0, |x_{lm}|- w_{lm},|z_{lm}|-\sqrt{u_{lm}v_{lm}}~).
\end{equation}
As we can clearly see from the above equation that the necessary condition for concurrence to be non-zero is the off-diagonal correlation function has to be greater than the diagonal function ,  either $|z_{lm}|> \sqrt{u_{lm}v_{lm}}$ or/and $|x_{lm}|>w_{lm} $. 
We find that the nearest-neighbour concurrence $C(1,2)=0$, as expected, for values of the interaction strengths and the magnetic field in the cluster model ground state. 
In fact, the off-diagonal correlation function for the nearest neighbours is zero for most of the parameter range,  though it is nonzero near the spin conserving  degeneracy point $J_y=J_x$ and $h=0$,  but not large enough to give a nonzero concurrence. 
\begin{figure*}[t]
\includegraphics[width=12cm]{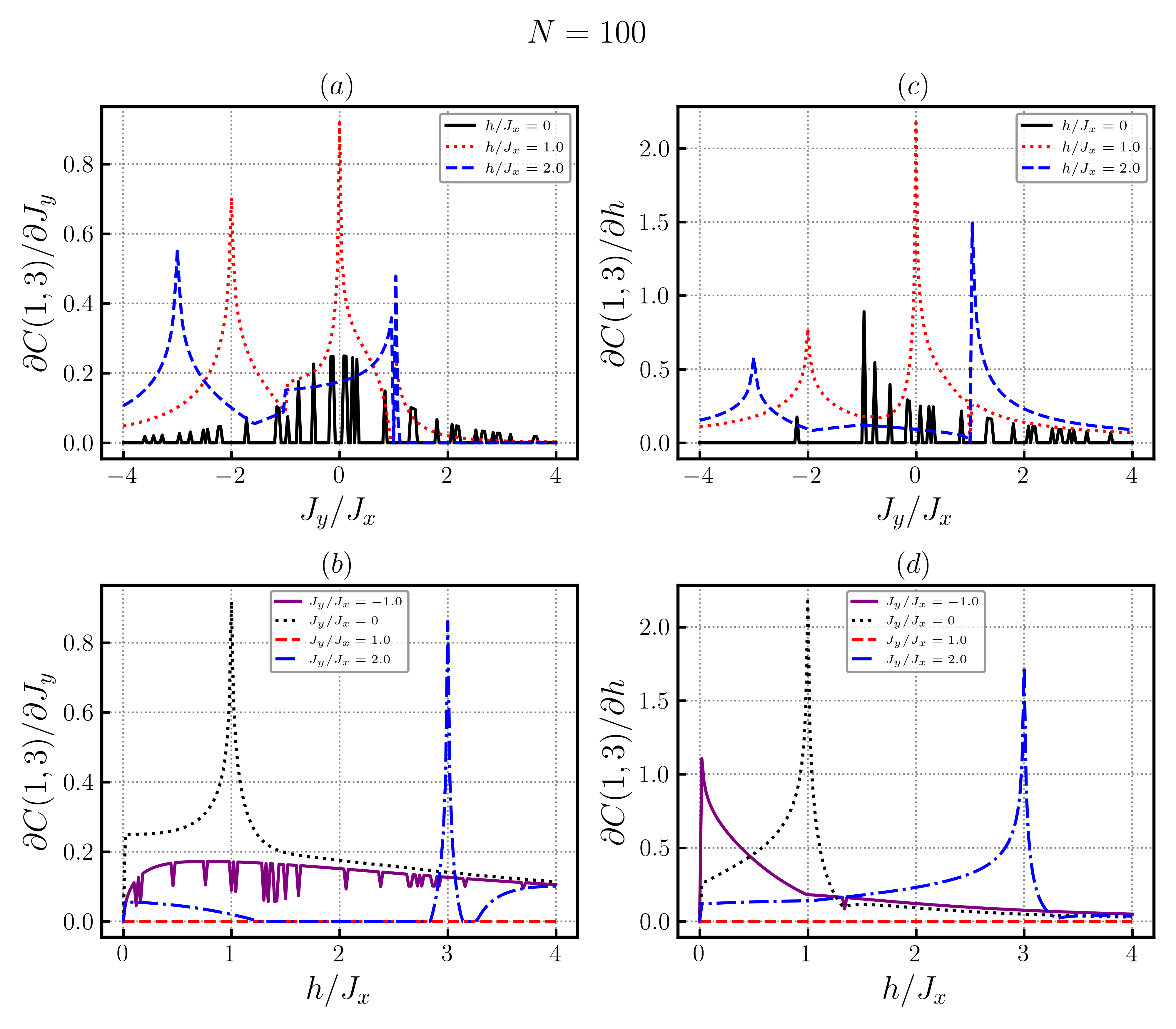}
\caption{ 
The first-order derivatives of the concurrence $C(1,3)$ are plotted against the ratio $J_y/J_x$ and $h/J_x$, displaying singular peak structure near degeneracy critical points, using $J_x=1$. The derivative $\partial C(1,3) / \partial J_y$ is plotted in (a) as a function of the ratio $J_y/J_x$  for different value of the magnetic field, and in (b) as a function of $h/J_x$ for different values of interaction strengths $J_y/J_x$ showing singular behaviour near degeneracy critical points. For $h=0$, there are a large number of degeneracy points as $J_y$ is varied, and similarly for $J_y=-1$ as $h$ is varied. 
 Similarly, the derivative  $\partial C(1,3) / \partial h$  plotted in (c)  as a function of $J_y/J_x$, and in (d) as a function of the magnetic field, shows singular peaks.  As seen from (b) and (d), both the derivatives are zero for the spin conserving case  $J_y=J_x$, for all values of $h$, though there is a line of degeneracy critical points.
 }
  \end{figure*}
The concurrence for distant pairs  is difficult to calculate analytically as matrix elements of reduced density matrix  involve more than four fermion operators.  Our numerical computation for finite systems reveals that the concurrence $C(1,4)$ and  $C(1,6)$ and so on are also zero for any value of magnetic field and interaction strengths. as the even and odd sites are decoupled. The concurrences $C(1,5)$ and  $C(1,7)$ are nonzero in the ground state, but  are substantially smaller than $C(1,3)$.

The pair concurrence depends on the both the diagonal correlation function and the off-diagonal correlation functions, as shown above. In particular $C(1,3)$ will
depend on the diagonal correlation function $\langle \sigma_1^z\sigma_3^z\rangle$ through the diagonal matrix elements $u_{13}, v_{13}, w_{13}$, and the off-diagonal correlation functions $\langle \sigma_1^+\sigma_3^+\rangle, \langle \sigma_1^+\sigma_3^-\rangle$ through the off-diagonal matrix elements $z_{13}$ and $x_{13}$. We have shown in Fig.2  the off-diagonal and diagonal correlations functions  for the next neighbour pair of spins plotted as functions of the ratio $J_y/J_x$ (top) for various fixed values of the magnetic field, as functions of the magnetic field (bottom) for various fixed values of $J_y/J_x$ for a finite chain of size $N=100$.  The off-diagonal correlation functions are anti-symmetric with the magnetic field, and the diagonal correlation function is symmetric with the field, thus the plots are shown for positive values of $h$ only. As a functions of $J_y/J_x$, the correlation functions shown in Fig.2(a), Fig.2(c) and Fig.2(e) show smooth behaviour for various values of the magnetic fields. 
The off-diagonal functions are shown also for $h/J_x=0.1$, as they vanish for $h=0$, as the interaction strength is varied.
The off-diagonal function $\langle \sigma_1^+\sigma_3^+\rangle$ is clearly zero for the spin conserving case $J_y/J_x=1$,  and the diagonal function shows a dip at the point.  As a function of $h$, the off-diagonal function $\langle \sigma_1^+\sigma_3^-\rangle$ shown in Fig.2(b), and the diagonal correlation function shown in Fig.2(e), show a strong signature of the line of degeneracy points for $J_y=J_x$, displaying a staircase behaviour similar to the magnetisation. The other off-diagonal correlation function shown in Fig.2(d) is clearly zero for all values of the magnetic field for the spin conserving case $J_y=J_x$. The main feature we  see here is that, for the spin non-conserving case, the off-diagonal function $\langle \sigma_1^+\sigma_3^+\rangle$ is substantially nonzero, whereas for the spin conserving case $J_y=J_x$, it is the off-diagonal function $\langle \sigma_1^+\sigma_3^-\rangle$ that is nonzero. We will see below, in either case, a nonzero concurrence is supported for some range of the magnetic field.

We now turn to the behaviour of $C(1,3)$,  the next-neighbour pair concurrence. In Fig.3, we have shown $C(1,3)$  in the ground state of the cluster spin model for various parameter values. Fig.3(a) shows the next neighbour concurrence as a function of $J_y/J_x$ for different values of the magnetic field.  For $h=0$, the concurrence $C(1,3)$ is zero for all values of the interaction strengths. For $h/J_x=1$, the concurrence is nonzero only when the interaction strengths have different sign, with a maximum near $J_y/J_x=1$, and vanishes with a singular behaviour near $J_y/J_x=0$ ( the degeneracy point exists for $h/J_x=1+J_y/J_x$).
We will investigate below the singular behaviour using the derivatives of the concurrence with the magnetic field and the interaction strength.  For $h/J_x=2$,  the concurrence shows singular behaviour for the spin conserving case of $J_y=J_x$. The concurrence has larger values  for $J_y/J_x<0$.  In Fig.3(b), the concurrence is shown as. a function of the magnetic field, for different values of $J_y/J_x$. As the concurrence is symmetric with $h$, only the positive $h$ range is shown. 
\begin{figure*}[t]
\includegraphics[width=\textwidth]{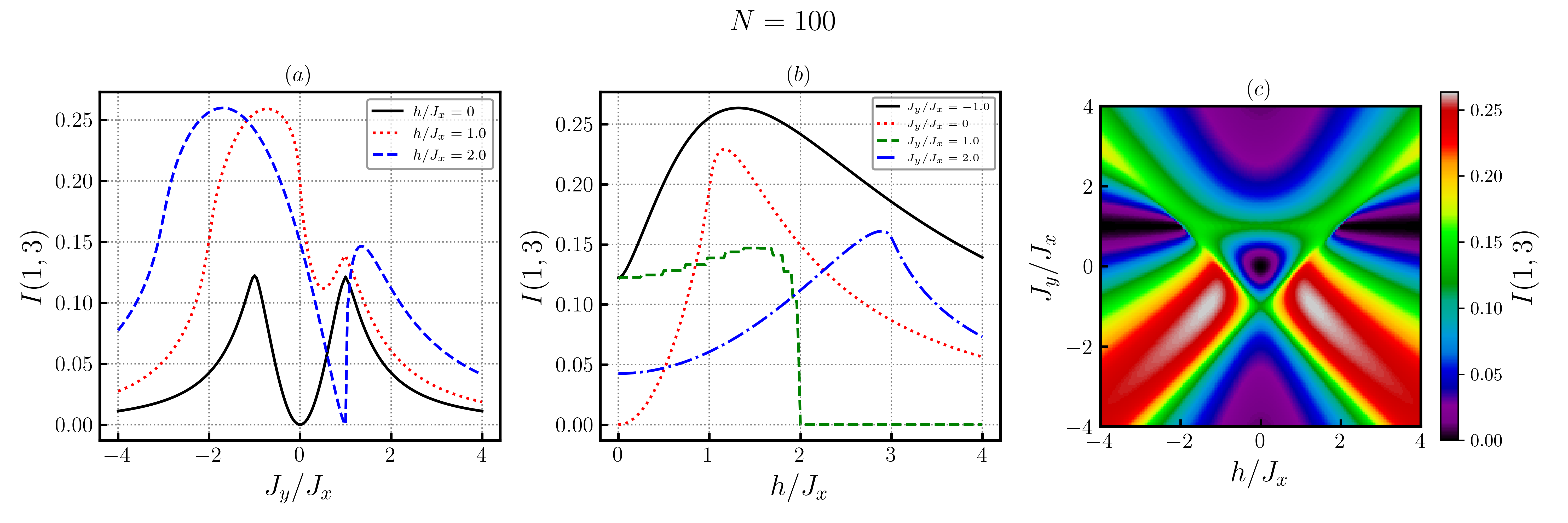}
\caption{ Quantum Mutual Information $I(1,3)$ of next-nearest neighbours, for $N=100$. (a) It is  plotted as a function of $J_y/J_x$, the ratio of interaction strengths for different values of magnetic field $h$. It shows peaks near degeneracy points, analogously as seen for the concurrence.
(b) $I(1,3)$ is plotted as a function of $h/J_x$.  For the spin conserving case, it shows a jagged structure, and zero beyond $|h/J_x|=2$. For other values of $J_y/J_x$, it does not show a cut-off behaviour.  (c) The density plot of $I(1,3)$ as a function of $h/J_x$ and $J_y/J_x$. Unlike the concurrence, it is non zero for a wide range of the parameters. $I(1,3)$ seems to take larger values for $J_y/J_x<0, |h/J_x|<2$. 
}
\end{figure*}
The concurrence becomes nonzero only after the field builds up to some value, depending on the interaction strength. We can see that the concurrence takes larger values for $J_y=-J_x$ compared to the other values of the interaction strengths.  For the spin conserving case $J_y=J_x$, it exhibits a staircase structure, due to the existence of a line of degeneracy points, analogous to the behaviour of the magnetisation, and the off-diagonal and diagonal correlation functions shown in Fig.2(b) and Fig.2(e) respectively. In this case, the concurrence is zero beyond an upper cutoff of $h/J_x=2$, as the ground state is a direct product state as we discussed in the last section.
Fig.3(c) shows the density plot of $C(1,3)$ as a function of the ratio of the  interaction strengths and the magnetic field. For $J_y=J_x$, and $h/J_x=\pm 2$, we can notice an elliptical island structure, with very small concurrence values. In this case, the off-diagonal correlation function $\langle \sigma_1^+\sigma_3^-\rangle$ is nonzero, and is strong enough to give a nonzero concurrence. Clearly, $C(1,3)$ has takes larger values of 0.2 for negative values of
$J_y/J_x$, and $|h/J_x|,2$.  Near the spin conserving point $J_y=J_x$, the concurrence seems to be quite small. 
Near the isolated degeneracy point $h=0, J_y=-J_x$, the concurrence is zero, though the off-diagonal correlation function $\langle \sigma_1^+\sigma_3^+\rangle$ is nonzero here, but not strong enough to make the concurrence nonzero.

The singular behaviour of  $C(1,3)$ can be seen more easily from its derivatives, with respect to the ratio $J_y/J_x$ and $h/J_x$. The derivatives $\partial C(1,3) / \partial J_y$  and $\partial C(1,3) / \partial h$ are expected to show singular peak structure as $h/J_x$ or $J_y/J_x$ are varied, similar to the nearest neighbour concurrence derivative in the transverse-XY model\cite{osterloh}. In Fig.4 we have shown the behaviour of the derivatives of $C(1,3)$ for the cluster model ground state. The derivative with respect to the ratio $J_y/J_x$ is shown in Fig. 4(a) as a function of $J_y$ for fixed values of the magnetic field $(h/J_x=0, 1., 2.0)$, exhibiting singular peaks corresponding to the degeneracy critical points. For $h=0$, there are a large number of peaks as $J_y/J_x$ is varied corresponding to the degeneracy points.  The $J_y-$derivative is plotted against the magnetic field for various values of $J_y/J_x$ in Fig.4(b). We have shown only for positive $h$, and a similar behaviour is seen for negative $h$. It shows a singular peak  for $J_y/J_x-0,2$, and several peaks for $J_y/J_x=-1$. Interestingly, for  the spin conserving case $J_y/J_x=1$, even though there is a line of degeneracy critical points as $h$ is varied, the derivatives are zero, However, for this case, $C(1,3)$ itself shows a singular staircase structure as shown in Fig.3(b). The derivative with respect to the magnetic field is shown in Fig.4(c) and FIg.4(d) as a function of $J_y/J_x$ and $h/J_x$ respectively, showing a similar singular peak structure. In this case also, for $J_y/J_x=1$, the $h-$derivative is zero for all values of the magnetic field.

\section{Quantum Mutual Information} \label{section:QMI}

The quantum mutual information is a measure of quantum correlations, widely used in the quantum information theory \cite{groisman}. It quantifies the amount of information shared between two subsystems, encompassing both classical as well as quantum correlations. The quantum mutual information $I(l,m)$  for a joint two-qubit state $\rho_{lm}$, for the two qubits at sites $l$ and $m$,  is defined using von Neumann entropies of the joint state, and the individual single-qubit density matrices $\rho_l$ and $\rho_m$, given by
\begin{equation}
I(l,m)= S(\rho_{l})+S(\rho_{m})- S(\rho_{lm}).
\end{equation}
Here, the von Neumann entropy $S(\rho)$ for a density matrix state $\rho$ is 
$S(\rho_{l})= -Tr(\rho_{l}\,log_{2}\,\rho_{l})$. For a pure two-qubit state, the quantum mutual information is twice the the entanglement between the two parties. Unlike the concurrence, which is calculated in a complicated way on the eigenvalues of the product of the joint density matrix and its time-reversed counterpart, the mutual information $I(l,m)$ depends only on the eigenvalues of the two-qubit joint state and the single-qubit marginals. The structure of the single-qubit density matrix is shown in Eq.12, which is already in the diagonal form. One eigenvalue of the single-qubit density matrix is $\langle (1+\sigma_l^z)/2\rangle= n_l=\gamma(0)$ (given in Eq.17), and the other eigenvalue is $1-\gamma(0)$. 
 The eigenvalues for the two-qubit reduced density matrix $\rho_{lm}$ , shown in Eq.15,  can be straightforwardly calculated as the density matrix has a block-diagonal form.
 The four eigenvalues of the two-qubit joint state are   
  $[(u_{lm}+v_{lm})\pm \sqrt{(u_{lm}-v_{lm})^2 + 4 |x_{lm}|^2}]/2 $ and $w_{lm}\pm  |z_{lm}|$. Thus, the quantum mutual information for nearest neighbour pairs and the next neighbour pairs can be calculated analytically, using the matrix elements  calculated in the last section.

For the cluster model ground state, $I(1,2)$ the quantum mutual information  for nearest neighbour pairs is zero, analogous with the concurrence discussed in the last section, for all parameter values of the magnetic field and the ratio of the interaction strengths, except for the spin conserving case $J_y=J_x$ and $h=0$ where it has a negligibly small value of about 0.001. 
The quantum mutual information $I(1,3)$ for next neighbours is shown in Fig.5  as a function of  $J_y/J_x$ and $h/J_x$ for $N=100$.
The  behaviour of $I(1,3)$ is not exactly similar the concurrence $C(1,3)$, as the mutual information includes both classical and quantum correlations.  In Fig.5(a), the next neighbour quantum mutual information $I(1,2)$ is plotted as a function of $J_y/J_x$ for particular values of the magnetic field, showing a peak near the degeneracy point.
In Fig.5(b), it is plotted as a function of the magnetic field for different values of $J_y/J_x$. Similar to the behaviour of $C(1,3)$, the quantum mutual information shows a jagged structure for the spin conserving case $J_y=J_x$ as the field is varied, corresponding to the line of degeneracy critical points. However, the quantum mutual information does not show a cut-off point in magnetic field, as seen in Fig. 3(b). The density plot of $I(1,3)$ is shown in FIg.5(c) as a function of both $J_y/J_x$ and $h/J_x$. $I(1,3)$ can be seen to have nonzero value almost for all parameter values, unlike vanishing $C(1,3)$  for a wide range of parameter values  as seen in FIg. 3(c). A dark patch, corresponding to near-zero quantum mutual information can be seen for $|h/J_x|>2$ in the spin conserving case of $J_y=J_x$, as expected for an uncorrelated ground state in this range.

\section{Conditional entropy and Quantum Discord}
 \begin{figure*}[t]
\includegraphics[width=\textwidth]{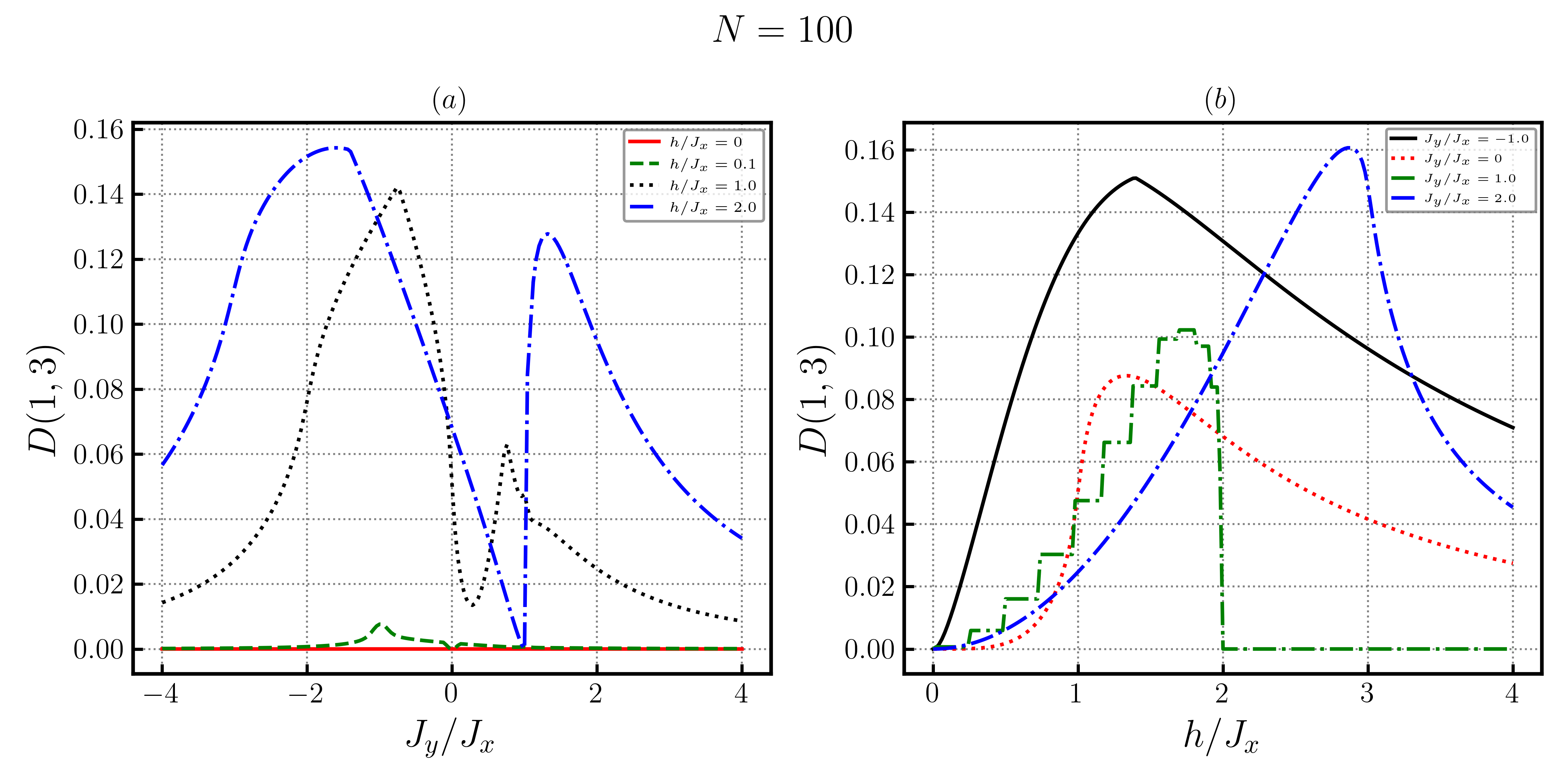}
\caption{ The quantum discord  $D(1,3)$ for next neighbour pair of spins for $N=100$.  (a) It is plotted against $J_y/J_x$, for different values of $h/J_x$. It shows a similar behaviour as the quantum mutual information.  For near zero field, the discord is nonzero only in the vicinity of $J_y/J_x=-1$. (b) It is plotted against $h/J_x$. Similar to the behaviour of the concurrence $C(1,3)$, the discord shows a staircase behaviour for the spin conserving case $J_y=J_x$, with a cut-off at $h/J_x=3$.
The discord $D(1,3)$ is zero for $h=0$  for all values of $J_y/J_x$, similar to the  concurrence, unlike the quantum mutual information. Surprisingly, the peak values here for
$J_y/J_x|\le 0$ are less than the corresponding values for the concurrence seen in Fig.3(b).
}
\end{figure*}

In analogy with the classical information theory, the bipartite  mutual information can be defined in an alternative way  using the conditional entropy. The conditional entropy of qubit $l$ after knowing the information of qubit $m$. In the quantum case, 
the conditional entropy uses the conditional density matrix $\rho_{l|m}$, that is the density matrix of $l$, after a measurement is performed on $m$ using a certain measurement basis for qubit $m$. Since there are two distinct outcomes for qubit $m$ after the measurement with different probabilities, the conditional entropy of $l$ conditioned on the measurement outcome of $m$, is a weighted sum of the von Neumann entropy of $\rho_{l|m}$.

 Let the measurement basis  be  $|\tilde 0\rangle$ and $|\tilde 1\rangle$ for qubit $m$, that are linear combinations of $\sigma^z$ basis states, given as $\ket{\tilde{0}} = \cos \frac{\theta}{2} \ket{0} + e^{i\phi} \sin{\frac{\theta}{2}}\ket{1},~
\ket{\tilde{1}} = \sin \frac{\theta}{2} \ket{0} - e^{i\phi} \cos{\frac{\theta}{2}}\ket{1}$.
After the measurement, the conditional state of qubit $l$ is either $\rho_{l|m_{\tilde 0}}$ with a probability $p_{\tilde 0}$  when qubit $m$ is in $|\tilde 0\rangle$, or  $\rho_{l|m_{\tilde 1}}$ with a probability $p_{\tilde 1}$, that correspondingly depends on the measurement outcome for qubit $m$, Thus, the conditional entropy $Q_{\theta,\phi}(l,m)$ will have measurement basis dependence, and is given by
\begin{equation}
Q_{\theta,\phi}(l,m) = p_{\tilde{0}}\,\, S(\rho_{l|m_{\tilde{0}}}) + p_{\tilde{1}}\,\, S(\rho_{l|m_{\tilde{1}}}).
\end{equation}

In analogy with classical information theory, we can define a quantum mutual information  $J(l,m)$ using the above conditional entropy, except that it depends on the measurement basis, as
\begin{equation}
J(l,m)= S(\rho_l)- Q_{\theta,\phi}(l,m),
\end{equation}
which measures the difference of the von Neumann entropy of qubit $l$ and its conditional entropy that is conditioned over the measurement outcome probabilities of the qubit $m$. In contrast, the quantum mutual information $I(l,m)$ shown in Eq.23 depends only on the individual von Neumann entropies of the qubits $l$ and $m$, and the von Neumann entropy of the joint state.  Using Bayes' relation between the conditional probability, the joint probability and
the marginals, $I(l, m)$ can be shown to be equal to $J(l,m)$ classically. However, for quantum systems there is a disparity between the two definitions of the mutual information, as $J(l,m)$ has dependence on the measurement basis.  The difference between them takes many values depending on the measurement basis. The quantum discord is defined as the minimum difference between them,  i.e. minimum of $ I(l,m) - J(l,m)$ (which is non-negative) over all possible measurement bases for the qubit $m$,  we have
\begin{equation}
D(l,m)= min_{(\theta,\phi)} [Q_{\theta,\phi}(\rho_{l|m})]+S(\rho_{m})-S(\rho_{lm}).
\end{equation} 

The conditional entropy has been calculated analytically\cite{kundu} for a two-qubit density matrix of the X-form given in Eq.15 for all values of $\theta$ and $\phi$.  In our case  it is seen that the conditional entropy is minimum for $\theta=\pi/2, \phi=0$. Using the results given in\cite{kundu} for the probabilities and the eigenvalues of the conditional density matrices, we can write the minimum conditional entropy in terms of the Shannon binary entropy, $H(x)=-x {\rm ln} x-(1-x) {\rm ln} (1-x)$, given as
\begin{equation}
Q_{{\pi\over2},0}(l,m)= H(\zeta_{lm}),
\end{equation}
where $\zeta_{lm}$ is given in terms of the matrix elements if the two-qubit reduced density matrix $\rho_{lm}$, as $\zeta_{lm}={1\over2} + \sqrt{({u_{lm}-v_{lm}\over 2})^2+(|x_{lm}|+|z_{lm}|)^2}$. Thus, the quantum discord can be calculated from the matrix elements of the one-qubit and two-qubit reduced density matrices shown in Eq.13 and Eq.16 respectively.

For the nearest neighbour pair of spins, we have seen that $I(1,2)$ is zero. This directly implies that the nearest-neighbour quantum discord is zero for all parameter values, as it is non-negative (i.e. $I(l,m)\ge J(l,m)$).
The quantum discord for next  neighbours $D(1,3)$ is plotted in Fig.6 as a function of the ratio $J_y/J_x$ and  $h/J_{x}$, for $N=100$. It shows a similar behaviour as the quantum mutual information. In Fig.6(a),  it is plotted against the ratio of the interaction strengths. For $h\sim 0$, the quantum discord is zero for all values of the ratio, except near $J_y/J_x\sim -1$. In Fig.6(b) it is shown as a function of the magnetic filed.
 The  $D(1,3)$ is seen to be similar to the concurrence in some regions as $h$ varies, with a smooth peak  in the vicinity of the degeneracy critical point $h=1+J_y$. For the spin conserving case $J_y=J_x$,  the discord shows a staircase structure with a cut-off at $h/J_x=2$.
Unlike the  concurrence $C(1,3)$,  the quantum discord $D(1,3)$ does not show a cut-off value for the spin non conserving case  $J_y/J_x \ne 1$, implying that the residual quantum correlations seem be to be nonzero for all values of the magnetic field.  Though both the conditional entropy $Q_{(\pi/2,0)}$ and the single qubit entropy $S(\rho_{l})$ are maximum at $h/J_x=0$, but  the quantum discord at $h/J_x=0$ is zero, due to the negative contribution of the joint state entropy. The conditional entropy and the single qubit entropy $S(\rho_{l})$ both keep growing and saturate to a maximum value, as the magnetic field is increased,  in contrast to the quantum discord $D(1,3)$. which exhibits a peak in the vicinity of a degeneracy critical point,  and then gradually decreases as the joint entropy $S(\rho_{lm})$  pulls it down. 
 It is  surprising that, for $ J_y/J_x\le 0$, the maximum value of concurrence $C(1,3)$ is larger  than the maximum value of quantum discord $D(1,3)$ (as seen from Fig. 3(b) and Fig. 6(b); it is normally expected that the concurrence is less than the quantum discord in spin chains. However, for  $ J_y/J_x >1$, the peak value of the discord $D(1,3)$ is larger than that of $C(1,3)$.

\section{Global Entanglement}
In the previous sections, we examined the correlation measures such as the concurrence, the quantum mutual information, and the quantum discord for next neighbour pair of qubits, that quantify the local entanglement and the local  quantum correlations. The analytic methods get very complicated  beyond the next neighbour pair correlations. 
In this section, we will  investigate the multipartite entanglement behaviour. The global entanglement is suitable for this case and gives an insight into the multi-qubit entanglement structure. 

\begin{figure}[t]
\includegraphics[width=8cm]{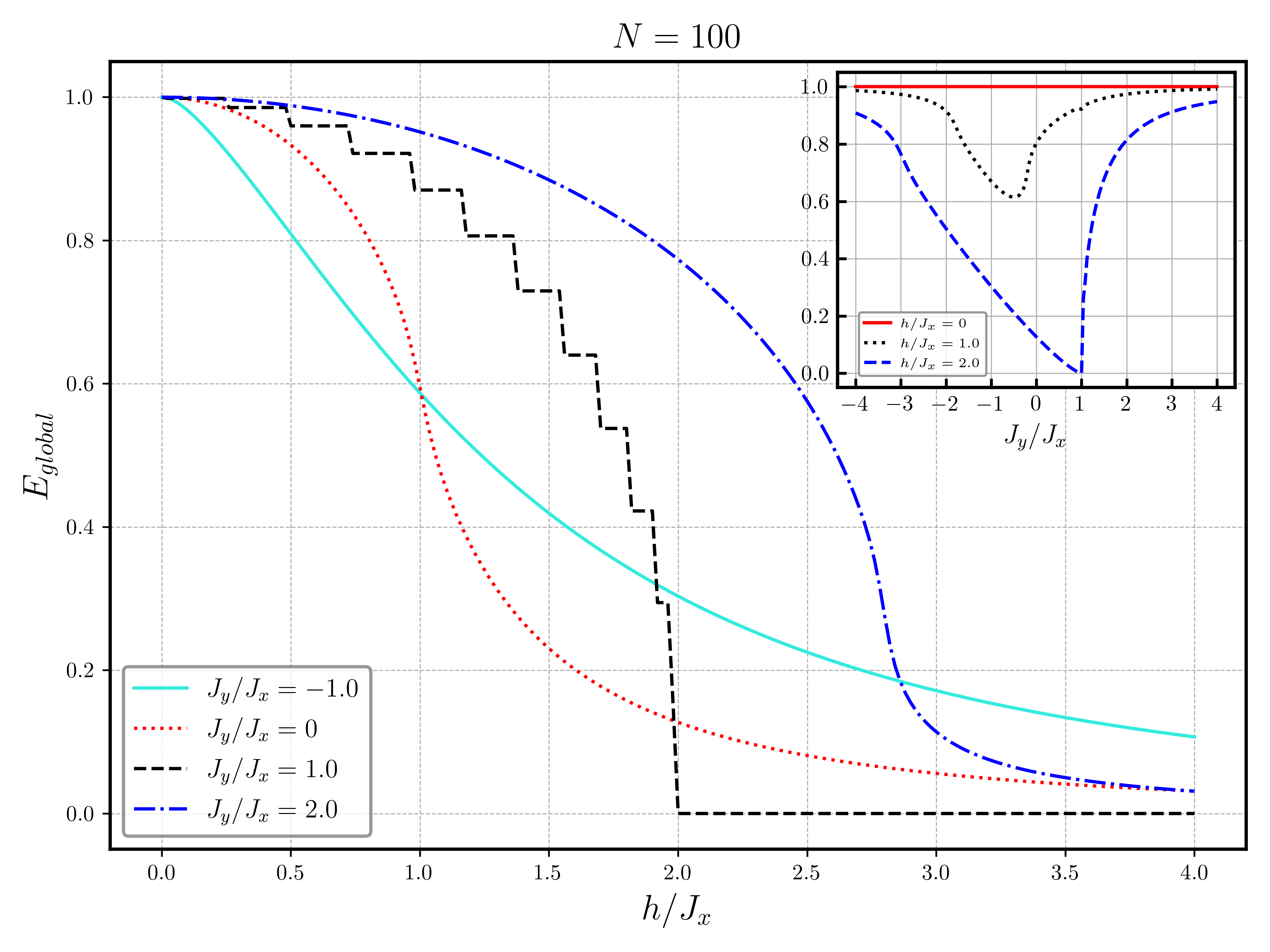}
\caption{
The global entanglement as a function of $h/J_x$ for various values of $J_y/J_x$, for $N=100$. For any value of $J_y$ it is maximum at $h=0$,  except for the spin conserving case $J_y=J_x$ where it exhibits a staircase structure.  The inset shows $E_{global}$ as a function of $J_y/J_x$ for different values of $h/J_x$. It shows a saturation for large values of $|J_y/J_x|$, for all values of the magnetic field.
}
\end{figure}
Meyer and Wallach introduced a measure for entanglement involving multipartite states\cite{Meyer}. This measure is formulated as a formal function applicable to many qubit pure states for $N$ number of spins, which has a straightforward physical interpretation. This entanglement measure behaves as an entanglement monotone, meaning it is a function that does not increase under local operations and classical communications. Brennen simplified the measure by utilising the invariance under local operations \cite{Brennen},   writing in terms of the individual one-qubit density matrices, given by
\begin{equation} \label{GE:GE1}
E_{global} = 2 \,\left( 1- \frac{1}{N} \sum_{j=0}^{N-1} \, Tr\, \rho^2_{j} \right).
\end{equation}
Here $\rho_{j}$ is the reduced density matrix corresponding to the qubit at the $j^{th}$ site obtained by partial trace over all the other sites of the ground state and $Tr[\rho^2_{j}]$ is quantifying the purity of the state $\rho_{j}$. The averaging procedure over all the spins, makes it a global entanglement measure.
From the translational invariance, all sites are equivalent; therefore, the global entanglement is just the average entanglement of a single site with the rest of the system. In this case, the single-site reduced density matrix is the same for all qubits,  represented by $\rho_{l}$ shown in Eq.12. The global entanglement can be written in a simple form,  we have

\begin{equation}
E_{global} = 4\left\langle c^{\dagger}_{j} c_{j}\right\rangle \left(1-\left\langle c^{\dagger}_{j} c_{j}\right\rangle \right).
\end{equation}
Thus, the global entanglement depends on the occupation probability of the $k$ states through Eq.13, similar to the magnetisation.

The behaviour of the  global entanglement $E_{Global}$ is shown in Fig.7  as a function of $h/J_x$ for different values of $J_y/J_x$, for a chain with $N=100$ spins.
As the magnetic field increases, the global entanglement smoothly decreases to zero; however, the range of $h$ increases as $J_y/J_x$ increases.
For the spin conserving case,  $J_x=J_y=1$, it exhibits a staircase structure due to the existence of a line of degeneracy critical points, similar to the behaviour of local quantum correlation measures. For the spin non-conserving case, it does not show different behaviour in the vicinity of degeneracy points. In the inset of  Fig.7,  we have shown the global entanglement as a function of $J_{y}/J_x$ for different  values of the magnetic field.  Similar to the other measures, the global entanglement shows different behaviour for $J_y/J_x$ negative and positive cases. For all values of  $J_y$, the global entanglement $E_{Global}$ takes the maximum value for $h/J_x=0$.
 For $h=0$, the fermion number operator $n_l$  shown in Eq.13 becomes independent of interaction strengths and remains constant at $1/2$, which results in the global entanglement $E_{Global}$ being unit for all values of $J_y$. This  behaviour is in contrast with the behaviour of the next neighbour pair concurrence and the quantum discord both of which vanish for zero magnetic field.  For all values of $h$, the global entanglement $E_{Global}$ approaches unity for large values of  $|J_y/J_x|$.

\section{Conclusions}

We have studied  several measures of quantum correlations for the ground state of the cluster spin model with three-spin interactions $J_x$ and $J_y$ along with a  transverse magnetic field $h$.
Every set of three neighbouring spins interact through XZX and YZY components of the respective spins. Spin chains with pairwise interactions, viz. the transverse-XY spin chain, exhibit only nearest neighbour quantum correlations in the ground state. In contrast, the cluster spin model studied here, exhibits  nonzero measures of quantum correlations only for the next neighbour pair of spins. In the case  $J_y=0, h=0$  
the ground state is the cluster state\cite{Shor}, a highly entangled state, that is used for measurement-based quantum computing\cite{briegel}. There are recent proposal for experimental realisation of three-spin interactions in optical lattices of ultra cold atoms\cite{pachos}. 

Though it has three-body interactions,  all the eigenstates can be exactly found using the Jordan-Wigner transformation, analogously as in the case of the transverse-XY   spin chain. The total z-component of the spin is conserved for $J_y=J_x$, and for all other values the dynamics is spin non-conserving, similar to the case of transverse-XY spin chain. For the spin conserving case,  for $|h/J_x|>2$ the ground state is a direct product state, either all spins up or down. Even with complicated three spin interactions, it is interesting to note that the ground state is an uncorrelated state for a range of parameters. It is the off-diagonal correlation function $ \langle \sigma_i^+\sigma_j^-\rangle$ that supports nonzero quantum correlation measures in the spin conserving case, and for the spin non-conserving case it is $\langle \sigma_i^+\sigma_j^+\rangle$ that supports quantum correlations.  There are many degeneracy critical points in the parameter space, with a degenerate ground state. The behaviour of the concurrence and the mutual information is different in the vicinity of degeneracy points. 

The  two-spin correlation functions, and various measures of quantum correlations are calculated analytically for the ground state of the cluster model. In the Jordan-Wigner fermion picture, the spin chain becomes union of two separate transverse-XY chains containing even-only sites, and odd-only sites, containing only next neighbour interactions for the spin chain.  Consequently, the nearest-neighbour quantum correlation measures, the concurrence, the quantum mutual information and the quantum discard are all zero.
The magnetisation, the concurrence, the quantum mutual information and the discord for the next neighbour pair of spins show different behaviour near the degeneracy points.  The next neighbour concurrence $C(1,3)$  exhibits singular  behaviour in the vicinity of the degeneracy critical points for the spin non-conserving case $J_y\ne J_x$.  However, in the spin conserving case,  $J_y-$ and $h-$ derivatives do not show singular peaks, implying that  the nature of degeneracy critical points  is different here. The global entanglement measure also shows a distinct behaviour near degeneracy points, exhibiting a singular behaviour, viz. a staircase structure, for the spin conserving case. Surprisingly, for $J_y/J_x\le 0$, the concurrence $C(1,3)$ has larger peak value than the peak value of the quantum discord $D(1,3)$.

\underline{Acknowledgement:} We would like to acknowledge  SERB, the science and engineering research board (India) for funding the research reported here, under  CRG scheme.


\end{document}